\documentclass[useAMS,usenatbib,usegraphicx]{mn2e}

\usepackage{aas_macros}

\usepackage{graphicx}

\title[Can filamentary accretion explain the MW satellites?]{Can filamentary accretion explain the orbital poles of the Milky Way satellites?}
\author[Pawlowski et al.]{M. S. Pawlowski$^{1}$\thanks{E-mail:
mpawlow@astro.uni-bonn.de}, P. Kroupa$^{1}$, G. Angus$^{2}$, K. S. de Boer$^{1}$, B. Famaey$^{1,3}$, \and G. Hensler$^{4}$\\
$^{1}$Argelander Institute for Astronomy, University of Bonn, Auf dem H\"{u}gel 71, D-53121 Bonn, Germany\\
$^{2}$Astrophysics, Cosmology \& Gravity Centre, University of Cape Town, Private Bag X3, Rondebosch, 7700, South Africa\\
$^{3}$Observatoire Astronomique, Universit\'e de Strasbourg, CNRS UMR 7550, F-67000 Strasbourg, France\\
$^{4}$Institute of Astronomy, University of Vienna, T\"urkenschanzstr.~17, A-1180 Vienna, Austria\\
}

\begin{document}

\date{Accepted 2012 April 24.  Received 2012 March 30; in original form 2012 January 17}

\pagerange{\pageref{firstpage}--\pageref{lastpage}} \pubyear{2012}

\maketitle

\label{firstpage}

\begin{abstract}
Several scenarios have been suggested to explain the phase-space distribution of the Milky Way (MW) satellite galaxies in a disc of satellites (DoS).
To quantitatively compare these different possibilities, a new method analysing angular momentum directions in modelled data is presented. It determines how likely it is to find sets of angular momenta as concentrated and as close to a polar orientation as is observed for the MW satellite orbital poles. The method can be easily applied to orbital pole data from different models.
The observed distribution of satellite orbital poles is compared to published angular momentum directions of subhalos derived from six cosmological state-of-the-art simulations in the Aquarius project. This tests the possibility that filamentary accretion might be able to naturally explain the satellite orbits within the DoS.
For the most likely alignment of main halo and MW disc spin, the probability to reproduce the MW satellite orbital pole properties turns out to be less than 0.5 per cent in Aquarius models. Even an isotropic distribution of angular momenta has a higher likelihood to produce the observed distribution. The two Via Lactea cosmological simulations give results similar to the Aquarius simulations. Comparing instead with numerical models of galaxy-interactions gives a probability of up to 90 per cent for some models to draw the observed distribution of orbital poles from the angular momenta of tidal debris. This indicates that the formation as tidal dwarf galaxies in a single encounter is a viable, if not the only, process to explain the phase-space distribution of the MW satellite galaxies.
\end{abstract}


\section{Introduction}
\label{sect:intro}

The satellite galaxies of the Milky Way (MW) are distributed in a highly inclined plane around the MW disc \citep{Lynden-Bell1976, Kroupa2005}, termed the Disc of Satellites (DoS). This is true individually for the 11 'classical' satellites \citep{Metz2007} as well as for the fainter ones detected in the Sloan Digital Sky Survey \citep{Kroupa2010}. In addition, globular clusters of the MW categorized as young halo clusters populate the same plane, and streams of stars and gas show a preference to align with it, too, thus being evidence for a vast polar structure (VPOS) around the MW \citep{Pawlowski2012}.

This strong spatial anisotropy is supported by the motions of these satellite galaxies. Using proper motion measurements, \citet[][M08 hereafter]{Metz2008} have derived the orbital poles (direction of angular momenta) for eight satellite galaxies. They found a strong alignment of six orbital poles close to the normal vector to the DoS, indicating that it is rotationally supported. In addition, the Sculptor dwarf galaxy is counter-orbiting, but also within the DoS. Thus, seven of eight satellites with measured proper motions have aligned orbital axes.

Several attempts have been made to explain the aforementioned spatial and orbital anisotropy within the standard model of cosmology. Dwarf galaxies might be accreted in groups \citep{LiHelmi2008, DOnghiaLake2009}, but \citet{Metz2009} have shown that such groups, observed extra-Galactic dwarf galaxy associations, are far too extended to explain the thin DoS structure. Comparing the orbital energy of MW satellites with dark matter subhalos from the Via Lactea II simulation, \citet{Rocha2011} find a wide spread in satellite infall times. This is in conflict with a collective accretion of the satellites in a single group. A common infall of MW satellites as former satellite galaxies of the LMC is investigated by \citet{Nichols2011b}, who conclude that ''the extended disk-of-satellites cannot be explained by the dwarfs being bound to the LMC within the last two apogalacticons, and may have another origin.'' Therefore, the idea of a group of galaxies centred on the LMC which was accreted is strongly disfavoured.

\citet{Libeskind2009} demonstrate that it is in principle possible that dark matter (DM) dominated satellites are similarly aligned as the most luminous MW satellites. However, this was shown to be an unsatisfactory solution, since \citet{Kroupa2010} demonstrated that their results indicate that only 0.4 per cent of all existing cold dark matter (CDM) halos of MW mass would host a galaxy similar to the MW with a similar spatial distribution of satellites. Furthermore, \citet{Libeskind2009} only resolve very massive subhalos (over $2.6 \times 10^9 M_{\sun}$), which is inconsistent with the lower dynamical mass estimates of most MW satellites \citep{Mateo1998, Walker2007, Strigari2008}. The significant over-abundance of predicted bright satellites compared to observations has first been identified by \citet{Bovill2011} and termed the ''bright satellite problem''. In addition to this discrepancy in absolute numbers, \citet{Kroupa2010} have shown the inconsistency of the predicted mass-functions of luminous subhalos with that of observed MW satellite galaxies.

A series of logically incompatible conclusions have been reached by authors working in the $\Lambda$CDM framework attempting to explain the MW satellite population: \citet{Nichols2011} argue that the dSph satellites had to have fallen in at a redshift $z > 3$\ in order for their gas content to be removed by means of tidal and ram-pressure stripping, while \citet{Deason2011} argue that the same dSph satellites must have fallen in recently in order for them to form the DoS and they discuss an example of group infall at $z = 0.6$. Apart from these mutually exclusive results, it has transpired that the proper motions of dSph satellites exclude infall \citep{Angus2011}. From Via Lactea II data Hensler \& Petrov (2012, in preparation) model the subhalo system around the MW and demonstrate that all DM subhalos should experience star formation, do not suffer tidal disruption, but only ram-pressure gas loss, and survive as faint dSphs. The most important issues are that all satellite dSph galaxies with subhalo masses larger than $10^6 M_{\sun}$\ have present-day surface brightnesses detectable by SDSS data and are isotropically distributed.

In one of the most recent attempts at explaining the DoS within $\Lambda$CDM, \citet[][L11 hereafter]{Lovell2011} have reported that CDM simulations naturally lead to satellites coherently rotating in quasi-planar distributions, such as the satellites of the MW. They calculated the directions of orbital angular momenta of DM subhalos in six galactic high-resolution halos taken from the Aquarius project \citep{Springel2008}. These are CDM simulations of similar halo masses to the assumed DM halo of the MW. Six different models have been analysed at resolution level 2, called Aq-A2 to Aq-F2. For details on the cosmological simulations and the method of determining subhalo angular momenta, the reader is referred to L11.
In that paper, while using the results to interpret the MW data, a quantitative test showing whether or not the angular momenta of the subhalos could feasibly represent the distribution of MW satellites was not given. Here we make that test, in particular because the L11 results have been used to argue that the DoS is naturally explained within $\Lambda$CDM \citep[e.g.][]{Keller2012}.

Motivated by the DoS and the apparently coherent orbits of the satellite galaxies within it, an alternative scenario of their origin has been proposed \citep{Lynden-Bell1976, Kroupa2005, Metz2008, Kroupa2010}. They might be tidal dwarf galaxies (TDGs), formed from the tidal debris in an encounter between the early MW and another, still gas-rich (proto-) galaxy \citep{Zwicky1956}. Galaxy collisions can lead to perpendicularly oriented discs of debris, with aligned orbits as well as counter-orbiting material as demonstrated in \citet[][P11 hereafter]{Pawlowski2011}. Such encounters are observed even in the present epoch, for example in the interacting system VV 340 \citep{Armus2009}. TDGs are found to form both in simulated \citep{Wetzstein2007, BournaudDuc2006, Bournaud2008} and observed \citep{Mirabel1992, Hunsberger1996, Weilbacher2003} galaxy collisions. They seem to be long-lived objects \citep{Kroupa1997, Recchi2007, Galianni2010, Duc2011} and share the same properties as dwarf elliptical galaxies (Dabringhausen et al., in prep.). This makes them a fundamental addition, if not even an alternative, to cosmologically formed dwarf galaxies.
In fact, that TDGs may be the dominant satellite dwarf galaxy population should be a necessary outcome of $\Lambda$CDM structure formation \citep{Okazaki2000}.

Here, a new method is presented which determines how likely it is to find as strongly a clustered distribution of orbital angular momentum vectors similarly close to the equator of the MW as the MW satellite orbital poles, if the angular momenta of the satellite population were drawn from the distributions in given models.
This method is then applied to the angular momentum directions of $\Lambda$CDM subhalos as determined and published by L11, to those derived from the two Via Lactea simulations \citep{Diemand2007, Diemand2008} and to those of tidal debris as determined from models of galaxy interactions.
Section \ref{sect:analysis} describes the assumptions, the observed situation, the models and the method. In Section \ref{sect:results} the results are presented, followed by the conclusions in Section~\ref{sect:concl}.

\section{Analysis}
\label{sect:analysis}

\subsection{Assumed galactic disc orientation in a DM halo}
\label{sect:assumptions}

\begin{figure}
\centering
 \includegraphics[width=80mm]{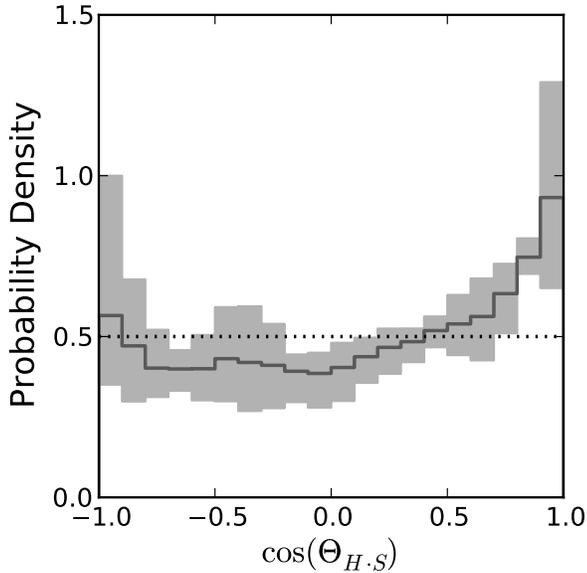}
 \caption{The data of figure 1 in L11, the cosine of the angle between the main halo spin and the subhalo angular momentum vector $\Theta_{H \cdot S}$, are plotted here as a histogram (input data courtesy of Mark R. Lovell).
For this plot, the subhalo orbital angular momenta of the six individual Aquarius simulations of L11 have been co-added. This results in the black, solid histogram, which gives the average probability density of the six simulations. The variation of the different simulations is illustrated by the grey, shaded area. For each bin, it illustrates the maximum and the minimum of the probability density determined for the six Aquarius halos individually.
Overall, the probability density of the simulations lies close to the isotropic case, illustrated by the horizontal dotted line. Co-rotating orbits are slightly preferred in all simulations. The deviation from the isotropic value is strongest for the bins closest to the DM spin axis, where the difference in probability density is up to a factor of two. Thus, while there is an over-abundance of subhalos with spins close to the main halo spin, all other spin directions are present in significant fractions. Note that the sum over all angles (of the area below the histogram) yields a probability of 1.
}
 \label{fig:lovell1}
\end{figure}

It is generally assumed that the angular momentum of the baryonic disc galaxy aligns with the angular momentum of the DM halo in which it resides. This assumption seems to be supported by the alignment of the minor axis of the inner part of DM halos with the axis of the disc galaxies in them \citep{Bailinetal2005}. \citet{Sharma2005} find a good correlation between baryonic gas and DM halo spin, with a mean misalingment-angle of only $18.9^\circ$ at a redshift of 0. The work of \citet{Bett2010} also shows that in simulations it is most likely for a galaxy to have its spin vector aligned with the direction of angular momentum of its parent DM halo. In particular the angular momentum of the inner halo (0.25 times the virial radius) aligns well with the galaxy spin, the median angle of misalignment being $23.9^\circ$. Taking the angular momentum of the whole halo, the median misalignment rises to $34.4^\circ$, but still the majority of galaxy spins are close to the main halo spins. Perpendicular orientations are rare (12 per cent within $\pm 15^\circ$ around the perpendicular orientation according to figure 17 in \citealt{Bett2010}), just as are dark halos in which the inner ($\leq 0.25\ r_{\rm{vir}}$) and outer regions are tilted by $\approx 90^{\circ}$.

In the following, it is therefore assumed that the galaxy rotates in the same direction as the parent halo, in line with what L11 state in their paper. It is assumed that the MW disc spin and the main halo spin in the simulations are parallel. In Section \ref{sect:differentequator} the requirement of this most likely alignment will be dropped, instead choosing an orientation of the disc galaxy which is most favourable to the formation of polar orbits, but puts the main halos spin and disc galaxy spin at an unlikely $90^\circ$-angle.

\subsection{The MW satellite orbital poles}
\label{sect:mpsatpoles}

\begin{figure}
\centering
 \includegraphics[width=80mm]{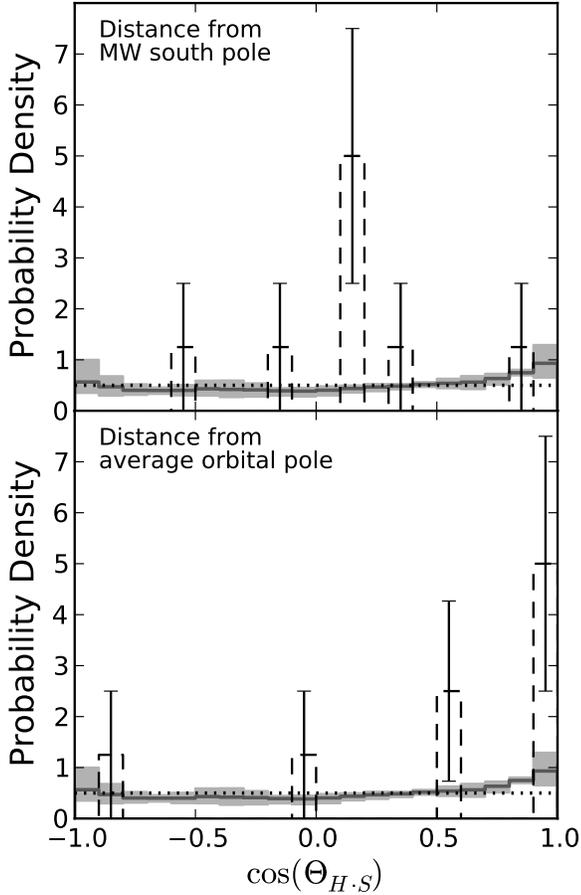}
 \caption{The same plot as shown in Figure \ref{fig:lovell1}, but note the different range on the Y-axis. In addition, this plot includes the orbital poles of the eight MW satellite galaxies (dashed black histograms with error bars) from M08, generated in the same way as the model histograms.
In the \textit{upper panel} it is assumed that the spin direction of the Milky Way galaxy, which points to the southern Galactic pole, aligns with the main halo spin, as L11 have argued. Therefore, the histogram shows the distance of the MW satellite orbital poles measured from the MW south pole. The distribution found in the simulations clearly differs from the observed one of the MW satellites. The MW satellites are preferentially on polar orbits, perpendicular to the MW disc spin and by implication to the spin of its supposed DM halo, at odds with the L11 results. Most of the eight known satellite orbital poles point close to the MW equator, about $90^\circ$\ away from the MW pole. They therefore appear close to the centre of the plot.
In the \textit{bottom panel} the most-favourable orientation of the spin of the DM halo of the MW is assumed. Here it is aligned with the average orbital pole of the MW satellite orbital poles from M08. Therefore, the histogram shows the distance of the MW satellite orbital poles measured from Galactic coordinates $(l, b) = (177.0^\circ,-9.4^\circ)$. The MW satellite orbital poles cluster much more than expected from the Aquarius simulations. Half of the poles are found in the bin closest to the assumed halo spin, whereas the simulations expect only about 10 per cent. All probability density distributions are normalised such that the sum over the area of all angular bins yields a probability of 1.
}
 \label{fig:lovell1b}
\end{figure}

The distribution of the MW satellite orbital poles (M08) shows two characteristic properties. First of all, the orbital poles cluster close to the normal to the MW DoS. The six best-aligned orbital poles show a spherical standard distance (M08) of only $\Delta_{\mathrm{sph}}^{\mathrm{MW}} = 35.4^\circ$. Secondly, the orbital poles preferentially fall close to the MW equator, the satellites thus move along on polar orbits. This is obvious from the direction of the average orbital pole for the six best-aligned orbital poles, it has an angular distance of only $d_{\rm{MW}} = 9.4^\circ$\ from the MW equator.

New proper-motion measurements for the satellite galaxies might lead to updated orbital pole directions, which in turn might change these values. The uncertainties of the orbital poles are the projected uncertainties of the angular momenta of the satellite galaxies. They are dominated by the, often large, uncertainties in the measured proper motions. M08 have shown that the uncertainties in the distance, position and radial velocity of satellite galaxies, the distance of the Sun from the Galactic centre and the circular velocity of the local standard of rest are negligible compared to the influence of the proper motion uncertainties on the directions of the orbital poles. But large proper-motion uncertainties do not necessarily result in large orbital-pole uncertainties. 
What is important for the determination of the orbital pole of a satellite galaxy is the direction of its motion with respect to its position relative to the Galactic centre, not its velocity.

As a result, the orbital poles of most satellites co-orbiting in the DoS are relatively well defined (directional uncertainties $< 15^\circ$). Only Draco and Carina show large uncertainties. However, their uncertainties fall on great-circles passing close to the average orbital pole of these six best-aligned orbital poles (see figure 1 of M08). This is why, on average, varying their orbital poles within the uncertainties cancels out when determining $\Delta_{\mathrm{sph}}^{\mathrm{MW}}$\ and $d_{\rm{MW}}$\footnote{Note that this is not inevitable. If the uncertainties would have been oriented perpendicular to this orientation, a variation of the satellite orbital poles along the uncertainties would preferentially lead to larger $\Delta_{\mathrm{sph}}^{\mathrm{MW}}$-values as the distance between the orbital poles would preferentially increase.}.
Within the current orbital pole uncertainties, updated proper motions can result in both larger or smaller values for $\Delta_{\mathrm{sph}}^{\mathrm{MW}}$\ and $d_{\rm{MW}}$, but the changes can be expected to be small. Therefore, in the following analysis the given values for the two parameters are adopted.

Figure \ref{fig:lovell1} is a remake of figure 1 in L11 using their data. It plots the distribution of the cosine of the angle between the main halo spin and the orbital angular momenta of individual subhalos, $\cos \Theta_{H \cdot S}$. All models peak close to $\cos \Theta_{H \cdot S} = 1$, so it is most likely for a subhalo to orbit in the same direction as the main halo. Three of the models also show an over-abundance of counter-orbiting subhalos close to $\cos \Theta_{H \cdot S} = -1$. In addition to the original data, in Figure \ref{fig:lovell1b} we have added the orbital poles of the MW satellites.
The upper panel assumes that the spin vector of the DM halo of the MW is parallel to the MW disc spin, pointing to the galactic south pole. The bottom panel assumes the halo spin coincides with the average orbital pole of the six best-aligned MW satellites. It constitutes the most favourable orientation of the halo to explain the preferred orbital direction of the MW satellites from the simulations.

In both cases, the observed distribution differs completely from the models. In the first case, the MW satellite orbital poles peak close to $\cos \Theta_{H \cdot S} = 0$, the MW equator. This is where the model distribution approaches its minimum, in contradiction to the observed distribution. In the second case, the observed orbital poles peak at the bin closest to the main halo spin by construction, which is also the maximum in the model distributions. But the peak contains 50 per cent of the orbital poles and is therefore much higher than all simulated models, which predict only about 10 per cent in this bin.

This illustrates that the observed situation does not follow the model. However, it does not rule out the \textit{possibility} that the observed orbital poles are drawn from one of the model distributions. The question we need a quantitative answer for is: \textit{How probable is this arrangement?}

\subsection{Models}
\label{sect:models}

This section describes the different models whose angular momentum direction distributions will be analysed using the method presented in the next Section. For that, it is necessary to repeatedly draw eight angular momentum directions randomly from their distributions.

\subsubsection{Isotropic distribution}
Assuming an isotropic distribution of angular momentum directions for satellite galaxies, 8 random directions are drawn from an uniform distribution on the sphere. These are used as angular momentum vectors.

\subsubsection{Aquarius Simulations}
To test the results of L11 derived from the Aquarius cosmological simulations, the angular momentum directions are drawn from the distribution of subhalo angular momenta\footnote{The angular momentum data for model Aq-B2 were provided in a different orientation to those shown in figure 4 in L11, therefore the precise position of the main halo spin is not known. It was estimated by rotating the provided data so that it visually resembles the orientation in figure 4 of L11. The distribution of angular momenta of this model in the re-production of their figure 1 also closely follows the original of figure 1 of L11, emphasising that the alignment is close to the correct one. Nevertheless, the results from this model are slightly more uncertain than those of the others. Fortunately, model Aq-B2 is the most isotropic one, so that this uncertainty has a negligible impact on the results.}. These are shown in their figure 4. Drawing from the whole population of subhalos and not only from the subhalos with the most massive progenitors should not pose a problem for this analysis, as L11 state that the latter trace the same structure of the whole subhalo sample.

\subsubsection{Via Lactea Simulations}
To provide a more complete picture of cosmological models, the subhalo angular momentum directions derived from the two Via Lactea simulations \citep[][D07 and D08 hereafter]{Diemand2007, Diemand2008} are analysed, too. These are cosmological n-body simulations of MW-sized DM halos with a quiet merger history. They are therefore possible hosts of a MW-type galaxy in a $\Lambda$CDM-universe. The analysis is based on the data freely available on the website of the Via Lactea project\footnote{http://www.ucolick.org/~diemand/vl/data.html} for the two models Via Lactea 1 (VL-1, D07) and Via Lactea 2 (VL-2, D08).
The subhalo angular momentum directions needed for the analysis are then determined for all subhalos in the available data set that are within the virial radius.

The halo spin is determined from the provided random sample of $10^5$\ particles at redshift zero. The sum of the angular momentum of all particles within $0.25~r_{\mathrm{vir}}$\ (VL-1: $r_{\mathrm{vir}} = 389\ \mathrm{kpc}$; VL-2: $r_{\mathrm{vir}} = 402\ \mathrm{kpc}$, see \citealt{Kuhlen2008}) is calculated, assuming each particle to have a mass such that the total mass of the simulation is reproduced. Then the sum of the subhalo angular momenta within the same radius is subtracted from the average particle angular momentum, which only leads to a small correction in the average spin direction of $1.8$\ (VL-1) and $11.4$\ (VL-2) degrees. The direction of the resulting vector is then adopted as the (inner) halo spin direction, which is more closely aligned with the disc galaxy spin as discussed in Sect. \ref{sect:assumptions}. This procedure is less sophisticated than the analysis L11 use, but can be easily done with the freely available data of the Via Lactea simulations and should give an estimate of the spin direction of the main DM halo.

\subsubsection{Tidal Models}
In addition to the cosmological simulations, models of a tidal origin for the MW satellites presented by P11 are analysed. In these models, a target disc galaxy collides with a perpendicularly oriented infalling disc galaxy on a polar orbit. P11 included equal-mass and 4-to-1 mass ratios for target-to-infalling galaxy. 
During the interaction, material is stripped off from the infalling galaxy and forms a disc of debris around the target galaxy, within which TDGs can form. Two populations of tidal debris form in most models, having pro- and retrograde orbits with respect to the orbit of the infalling galaxy. The models therefore not only produce orbits similar to the majority of those described by the satellite galaxy orbital poles, but can also account for the counter-orbiting direction of the Sculptor dwarf galaxy.

The distributions of angular momentum directions for these models are drawn from the tidal material, so from the angular momentum directions of individual particles. To show the evolution of the angular momentum distributions, three time-steps representing 5.0, 7.5 and 10.0 Gyr after the beginning of the model calculations are considered. These are chosen to be well after the initial perigalactic passage of the two galaxies and, in the merger case, also after the final collision that happens after 2.5 to 4 Gyr.
To include their angular momentum direction in the analysis, the particles are required to have a distance of at least 30 kpc from the central galaxy in order to avoid material from the galactic disc. If the model leads to a merger, particles with distances of up to 400 kpc are considered, while in a fly-by encounter particles can have distances from the target galaxy of up to half the distance between the two interacting galaxies.

The results of all models from P11 are compiled in the appendix. In the following, four models will be discussed in detail. The four particular models have not been chosen to give the best results, but to show the typical range of fractions determined for the tidal models of P11. They include two 1:1 mass ratio models with the infalling galaxy oriented prograde, one which results in a fly-by (''1:1-flyby-pro'', called 5deg200vel in P11) and one resulting in a merger (''1:1-merger-pro'', 7.5deg100vel in P11) of the two colliding galaxies. In addition, two 4:1 mass ratio merger models are included. One of these (''4:1-merger-pro'', 7.5deg100vel in P11), having a prograde infalling galaxy, is one of the models with a good agreement. The other model (''4:1-merger-retro'', 10deg100vel in P11) has an infalling galaxy in retrograde orientation and results in the worst agreement with the MW orbital poles of all tidal models. The reason is that, while the model forms a very long tidal tail of more than 300 kpc, it is dominated by a spheroid-like distribution of tidal particles out to about 50 kpc.

\subsection{Method}
\label{sect:method}

To assess how likely it is to find a similar distribution of angular momentum directions as inferred from the eight MW satellite galaxies, it has to be estimated how often similar parameters can be produced by a given model. For this, a sub-sample of 8 different angular momentum directions is drawn from the distribution of angular momentum directions in the model tested.

Of the 8 known orbital poles of the MW satellites, 6 are aligned with each other and the DoS normal direction. Analogously, for each possible combination of 6 out of the 8 angular momentum directions drawn from the models, the mean direction of the angular momenta is determined. Centred on these average directions, the spherical standard distance $\Delta_{\mathrm{sph}}$\ \citep{Metz2007} of the 6 corresponding angular momentum directions is calculated. It is defined as
$$
\Delta_{\mathrm{sph}} = \sqrt{ \frac{ \sum_{i=1}^{6} \left[ \arccos \left( \langle \hat{n} \rangle \cdot \hat{n}_{i} \right) \right] ^{2} }{ 6 } },
$$
where $\hat{n}_i$\ are the angular momentum direction unit vectors, $\langle \hat{n} \rangle$\ is the unit vector pointing into their mean direction and '$\cdot$' denotes the scalar product of the vectors. Note that the formula in \citet{Metz2007} deals with axial data and therefore has to take the absolute value of the scalar product, in contrast to our case.
$\Delta_{\mathrm{sph}}$\ is a measure for the clustering of orbital poles around their average direction. The sample of 6 angular momentum directions leading to the smallest $\Delta_{\mathrm{sph}}$\ is chosen. These are called the six best-aligned angular momenta from the sample of eight. The value for the average angular momentum direction and the spherical standard distance of these best-aligned angular momenta is stored. The spherical standard distance will be used for the clustering criterion in Sect. \ref{sect:clustering}.

Finally, the angular distance $d$\ of the average angular momentum direction of the six best-aligned angular momenta to the equator of the model is computed. In the isotropic case, the 'equator' is an arbitrary great circle as there is no preferred direction available. The great circle $90^{\circ}$\ away from the main halo spin is referred to as the 'equator' in the Aquarius and Via Lactea simulations, because the main halo spin aligns with the galactic disc spin, as was discussed in Sect. \ref{sect:assumptions}.  In the case of the galaxy-interaction models, the 'equator' is the plane of the target galaxy which resembles the orientation of the MW disc. The angular distance to the equator is used in the orientation criterion in Sect. \ref{sect:orientation}, assessing whether the average angular momentum hints at polar orbits or not. 

When this algorithm is applied to the eight orbital poles of satellite galaxies of the MW, the same six satellite galaxies as reported by M08 are found to have the best-aligned orbital poles. Consequently, the algorithm gives the same parameters as reported by M08: $\Delta_{\mathrm{sph}}^{\mathrm{MW}} = 35.4^\circ$\ and $d_{\rm{MW}} = 9.4^{\circ}$.

This process of drawing 8 angular momentum directions and determining the parameters $\Delta_{\mathrm{sph}}$\ and $d$\ for the best-aligned subsample of six of these is called one \textit{realisation}. To determine the statistical properties, $10^5$\ different realisations are produced for each model. The resulting distributions of the two parameters are shown in Figures \ref{fig:standarddistance} and \ref{fig:averagedistance}.

The method can be easily adjusted once more than eight satellite galaxy orbital poles become available through observations of proper motions. The number of angular momenta drawn for one realisation (currently eight) would need to be increased, the number of best-aligned poles (currently six) might be changed, and the parameters $\Delta_{\mathrm{sph}}^{\mathrm{MW}} = 35.4^\circ$\ and $d_{\rm{MW}} = 9.4^{\circ}$ adjusted to the observed situation. Note also that the analysis does not account for the alignment of the orbital poles with the DoS normal, nor does it factor in that the Sculptor satellite galaxy is counter-orbiting with respect to the six best-aligned orbital poles, but also orbiting in the DoS. If more satellite galaxies on counter-orbits are found, the analysis might need to be adjusted to account for this, e.g. by analysing not the directions of angular momenta, but the orbital axes (given by the angular momentum direction and its counter-direction), thus in effect combining co- and counter-orbiting poles on one half-sphere.

\begin{table*}
\begin{minipage}{180mm}
 \caption{Models and resulting fractions of realisations fulfilling the criteria}
 \label{tab:fractions}
 \begin{center}
 \begin{tabular}{@{}lrrrrrrrr}
  \hline
  Model & Reference & $N$ & $\bar{\Delta}_{\mathrm{sph}} [^\circ]$ & $\tilde{\Delta}_{\mathrm{sph}} [^\circ]$ & $f_{\Delta} [\%]$ & $f_{d} [\%]$ & $f_{\rm{both}} [\%]$ & $f_{\rm{indep}} [\%]$ \\
  \hline
Aquarius A2 & L11 & 30177    & 52.3 & 53.4 & $8.20 \pm 0.09$ & $6.00 \pm 0.08$ & $0.10 \pm 0.01$ & 0.49 \\
Aquarius B2 & L11 & 31050    & 53.3 & 53.8 & $4.00 \pm 0.06$ & $14.83 \pm 0.12$ & $0.44 \pm 0.02$ & 0.59 \\
Aquarius C2 & L11 & 24628    & 51.2 & 52.1 & $9.49 \pm 0.10$ & $9.28 \pm 0.10$ & $0.22 \pm 0.01$ & 0.88 \\
Aquarius D2 & L11 & 36006    & 53.4 & 54.2 & $4.53 \pm 0.07$ & $12.39 \pm 0.11$ & $0.40 \pm 0.02$ & 0.56 \\
Aquarius E2 & L11 & 30372    & 53.2 & 54.0 & $5.24 \pm 0.07$ & $9.68 \pm 0.10$ & $0.36 \pm 0.02$ & 0.51 \\
Aquarius F2 & L11 & 35041    & 52.1 & 52.7 & $5.40 \pm 0.07$ & $12.16 \pm 0.11$ & $0.44 \pm 0.02$ & 0.66 \\
Via Lactea 1 & D07 & 2576    & 53.2 & 53.9 & $4.26 \pm 0.07$ & $12.86 \pm 0.11$ & $0.41 \pm 0.02$ & 0.55 \\
Via Lactea 2 & D08 & 9381    & 53.1 & 54.1 & $6.62 \pm 0.08$ & $20.28 \pm 0.14$ & $1.49 \pm 0.02$ & 1.34 \\

Isotropic   & this paper & $\infty$ & 54.0 & 54.6 & $3.00 \pm 0.05$ & $16.41 \pm 0.13$ & $0.47 \pm 0.02$ & 0.49 \\

1:1-flyby-pro (5 Gyr) & P11 & 5821     & 31.3 & 7.8 & $60.13 \pm 0.25$ & $91.12 \pm 0.30$ & $60.11 \pm 0.25$ & 54.80 \\
1:1-flyby-pro (7.5 Gyr) & P11 & 5841     & 31.7 & 14.3 & $59.06 \pm 0.24$ & $86.79 \pm 0.29$ & $58.48 \pm 0.24$ & 51.25 \\
1:1-flyby-pro (10 Gyr) & P11 & 5756     & 31.6 & 20.4 & $58.39 \pm 0.24$ & $84.08 \pm 0.29$ & $57.27 \pm 0.24$ & 49.09 \\

1:1-merger-pro (5 Gyr) & P11 & 36438     & 22.4 & 9.9 & $74.36 \pm 0.27$ & $59.10 \pm 0.24$ & $48.40 \pm 0.22$ & 43.94 \\
1:1-merger-pro (7.5 Gyr) & P11 & 35954     & 21.5 & 12.1 & $79.27 \pm 0.28$ & $15.06 \pm 0.12$ & $8.96 \pm 0.09$ & 11.94 \\
1:1-merger-pro (10 Gyr) & P11 & 37302     & 22.8 & 12.2 & $77.42 \pm 0.28$ & $20.77 \pm 0.14$ & $12.03 \pm 0.11$ & 16.08 \\

4:1-merger-pro (5 Gyr) & P11 & 93940     & 17.8 & 8.7 & $82.30 \pm 0.29$ & $82.84 \pm 0.29$ & $74.69 \pm 0.27$ & 68.18 \\
4:1-merger-pro (7.5 Gyr) & P11 & 80454     & 18.4 & 14.2 & $88.73 \pm 0.30$ & $74.52 \pm 0.27$ & $70.58 \pm 0.27$ & 66.12 \\
4:1-merger-pro (10 Gyr) & P11 & 77768     & 25.4 & 22.8 & $73.89 \pm 0.27$ & $53.22 \pm 0.23$ & $46.45 \pm 0.22$ & 39.33 \\

4:1-merger-retro (5 Gyr) & P11 & 75431     & 46.2 & 49.3 & $33.58 \pm 0.18$ & $36.17 \pm 0.19$ & $19.22 \pm 0.14$ & 12.15 \\
4:1-merger-retro (7.5 Gyr) & P11 & 60248     & 47.8 & 49.1 & $16.20 \pm 0.13$ & $28.92 \pm 0.17$ & $6.97 \pm 0.08$ & 4.69 \\
4:1-merger-retro (10 Gyr) & P11 & 61157     & 50.4 & 51.1 & $8.11 \pm 0.09$ & $13.17 \pm 0.11$ & $1.04 \pm 0.03$ & 1.07 \\

  \hline
 \end{tabular}
 \end{center}

 \small \smallskip

Model: Name of the model or simulation.
Reference: The original publication presenting the respective model (L11: \citealt{Lovell2011}; D07: \citealt{Diemand2007}; D08: \citealt{Diemand2008}; P11: \citealt{Pawlowski2011}).
$N$: Number of angular momentum directions to draw realisations from.
$\bar{\Delta}_{\mathrm{sph}}$: Mean of the spherical standard distance distribution.
$\tilde{\Delta}_{\mathrm{sph}}$: Median of the spherical standard distance distribution.
$f_{\Delta}$: Fraction of realisations fulfilling the clustering criterion (having a spherical standard distance of no more then $35.4^\circ$, the value for the MW satellites). 
$f_{d}$: Fraction of realisations fulfilling the direction criterion (having an average angular momentum direction pointing no further away from the equator than $9.4^\circ$, the value for the MW satellites).
$f_{\rm{both}}$: Fraction of realisations fulfilling both criteria.
$f_{\rm{indep}} = f_{\Delta} \cdot f_{d}$: Fraction fulfilling both criteria if they were independent.
All uncertainties are estimated assuming Poisson statistics.
\end{minipage}
\end{table*}

\section{Results}
\label{sect:results}

The results are listed in Table \ref{tab:fractions}, together with the total numbers of individual angular momentum directions of each model. In addition and for completeness, the results for all tidal models of P11 are compiled in the Appendix.

\subsection{Fulfilling the clustering criterion}
\label{sect:clustering}

\begin{figure*}
\centering
 \includegraphics[width=180mm]{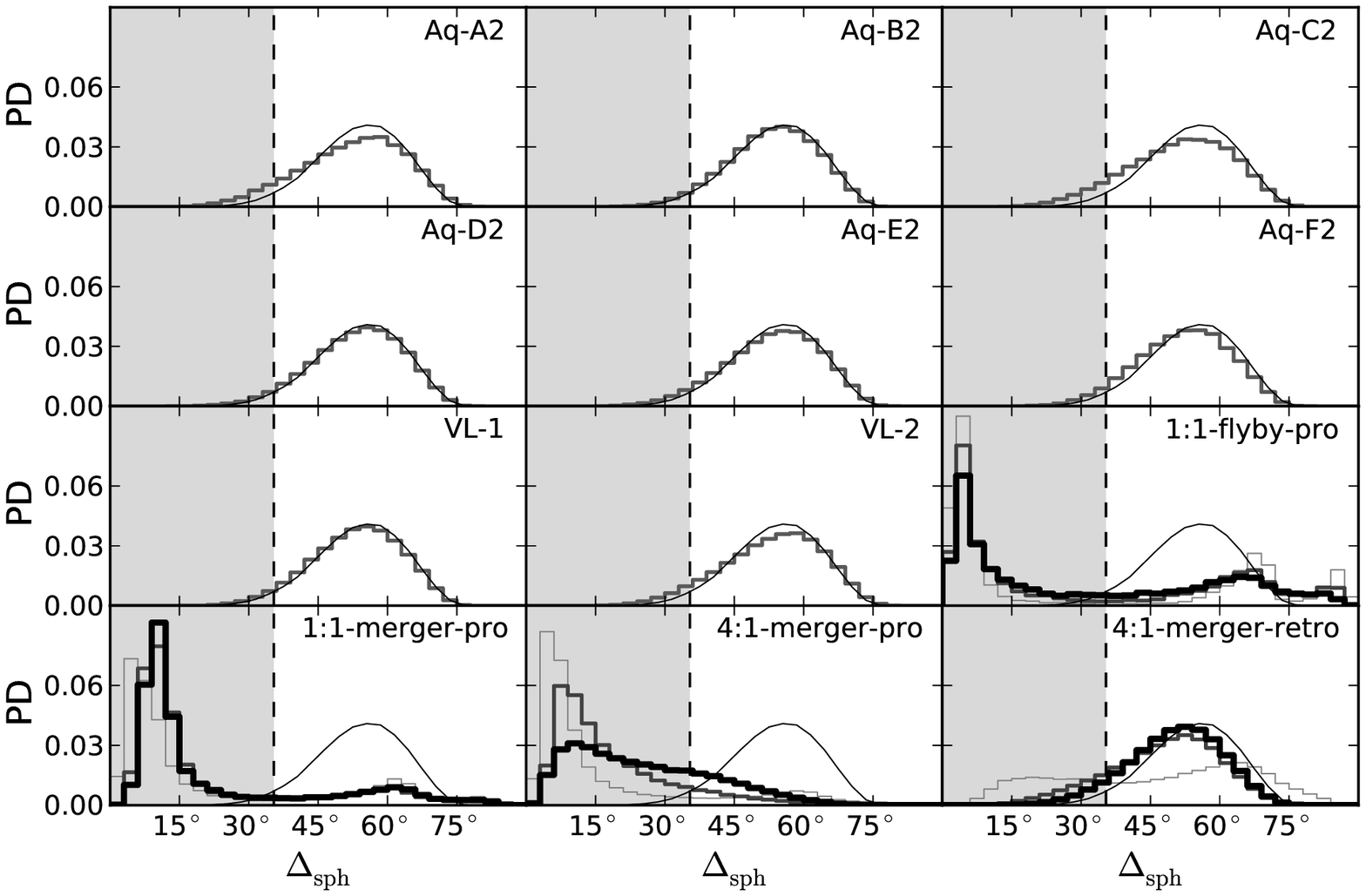}
 \caption{Distributions of the probability density (PD) of the spherical standard distance $\Delta_{\mathrm{sph}}$ of the 6 most closely aligned angular momentum vectors out of 8 different, randomly drawn angular momentum vectors. In the first eight panels, the angular momentum vectors are drawn from the subhalo angular momentum vectors of the Aquarius simulations presented by L11 and from the Via Lactea simulations by D07 and D08. The last four panels show the distribution that results when drawing angular momentum vectors from the particles in calculations of galaxy interactions by P11. They each contain three distributions showing the time-evolution of each model. The thin, light gray histogram illustrates the situation at 5 Gyr after the start of the calculation, the dark gray histogram at 7.5 Gyr and the thick black histogram at 10 Gyr.
The case of an isotropic sample of angular momenta is included as the thin black line in all panels for comparison.
 The spherical standard distance for the six best-fitting orbital poles in the MW, $35.4^\circ$, is indicated by the vertical dashed lines. To fulfil the clustering criterion, a realisation has to fall to the left of this line, this allowed region being highlighted by a shading.
 The distributions resulting from the cosmological simulations are nearly indistinguishable from the isotropic case. The vast majority of determined spherical standard distances is significantly larger than the observed one. This is completely different for the most examples of tidal material, which show very strongly concentrated distributions of angular momentum directions, in agreement with the distribution inferred from the MW satellite galaxies.
 }
 \label{fig:standarddistance}
\end{figure*}

Figure \ref{fig:standarddistance} plots the resulting distributions of the spherical standard distances $\Delta_{\mathrm{sph}}$\ for the analysed models. The isotropic distribution peaks at about $55^\circ$\ and is spread between $30^\circ$\ and $75^\circ$. 

The six Aquarius and the two Via Lactea simulations lead to $\Delta_{\mathrm{sph}}$-distributions that are nearly indistinguishable from the isotropic distribution. The mean value of the spherical standard distances, $\bar{\Delta}_{\mathrm{sph}}$, are slightly lower, by up to about $3^\circ$\ compared to the isotropic case having $\bar{\Delta}_{\mathrm{sph}} = 54^\circ$.

The first three tidal models, in contrast, show a completely different behaviour in their $\Delta_{\mathrm{sph}}$-distributions. They display a very strong peak at 5 to $15^\circ$\ in all three time steps. This peak gets lower for later time steps in all three models, showing that the angular momentum distributions widen with time.
Their mean values $\bar{\Delta}_{\mathrm{sph}}$\ stay relatively constant at $31^\circ$\ (1:1-flyby-pro), $22^\circ$\ (1:1-merger-pro) and $18^\circ - 25^\circ$\ (4:1-merger-pro). This shows that the tidal models result in much more concentrated distributions of orbital poles than the Aquarius and Via Lactea simulations. They are qualitatively very different from the isotropic case and the concentration is stable for at least half a Hubble time. The $\Delta_{\mathrm{sph}}$-distributions show tails spreading out to almost $90^\circ$. These arise from the fact that the angular momenta of the tidal debris cluster in two opposite directions, both being close to the central galaxy's equator. In addition to the strong clustering of orbital angular momenta, this pro- and retrograde tidal debris can therefore explain the counter-orbiting direction of the Sculptor dwarf galaxy. However, this additional aspect of the tidal models being consistent with the observed situation is not considered by the present analysis.

The $\Delta_{\mathrm{sph}}$-distribution of the last tidal model (4:1-merger-retro) approaches an isotropic-like shape for the last time steps. This illustrates that the spheroid-like distribution of the merger remnant is dominating the angular momentum directions. On average, the distribution is only minimally more concentrated than those of the cosmological models.

The orbital directions of satellite galaxies can be changed by a number of processes, like scattering with other satellite galaxies \citep[which might explain the energetic orbit of Sagittarius, see][]{Zhao1998}, precession due to non-axisymmetric potentials or tidal torques from neighbouring galaxies. For a clustered distribution of orbital poles, these processes lead to an increase of the spherical standard distances. Therefore, the observed $\Delta_{\mathrm{sph}}$-value of the MW satellites can be interpreted as an upper limit. It might have been smaller in the past, but not larger.

To fulfil the \textit{clustering criterion}, a realisation has to be at least as well clustered as the MW satellite orbital poles. The $\Delta_{\mathrm{sph}}$-value of the MW satellites, $35.4^\circ$, is illustrated in Figure \ref{fig:standarddistance} by the vertical dashed line. All realisations with $\Delta_{\mathrm{sph}} \leq 35.4^\circ$\ are counted as passing the clustering criterion. Their number divided by the total number of realisations per model is $f_{\Delta}$, the fraction of realisations that fulfil the clustering criterion. The values of $f_{\Delta}$\ are compiled in Table~\ref{tab:fractions} for all models shown in the plot.

In the isotropic case, only 3 per cent of the realisations fulfil the clustering criterion. As can be expected from the longer tail towards lower $\Delta_{\mathrm{sph}}$-values in the Aquarius simulations, the fractions of realisations passing the clustering criterion are higher. However, this is not a strong effect, the increase compared to the value of isotropy is only a factor of $1.3$\ to $3.2$. It might thus be concluded that filamentary accretion can account for a minor increase in the concentration of angular momenta distributions of subhalos. The Via Lactea simulations lead to clustering results similar to those of the Aquarius simulations.

Strongly clustered distributions of angular momentum directions arise naturally in tidal interactions. The first three tidal models easily fulfil the clustering criterion, for each time step more than half of all realisations have $\Delta_{\mathrm{sph}} \leq 35.4^\circ$: about 60 per cent in model 1:1-flyby-pro, 75 per cent in 1:1-merger-pro and 80 per cent in the 4:1-merger-pro. There is only a small variation with time: $f_{\Delta}$\ tends to become lower with time (also compare the averages for all P11-models in tables \ref{tab:tidal4t1} and \ref{tab:tidal1t1}).

The last tidal model (4:1-merger-retro) starts with $f_{\Delta}$\ of 34 per cent at 5 Gyr, but this value drops to only 8 per cent for the last time step, where it is comparable to $f_{\Delta}$\ of the cosmological models. This shows that the initially clustered distribution of angular momentum directions in this model quickly disperses because the spheroid-like component of the merger remnant dominates over the tidal tail material. If only the tidal tail particles are included in the analysis by demanding a minimum distance from the galactic centre of 60 kpc, $f_{\Delta}$\ becomes 28 per cent for the final time step. This fraction is still relatively small compared to that of the other tidal models.

The clustering criterion is completely independent of the orientation of the main galaxy. It alone already shows that reproducing the orbital-pole-distribution of the MW satellites in the Aquarius and Via Lactea simulations is very unlikely.

\subsection{Fulfilling the orientation criterion}
\label{sect:orientation}

\begin{figure*}
\centering
 \includegraphics[width=180mm]{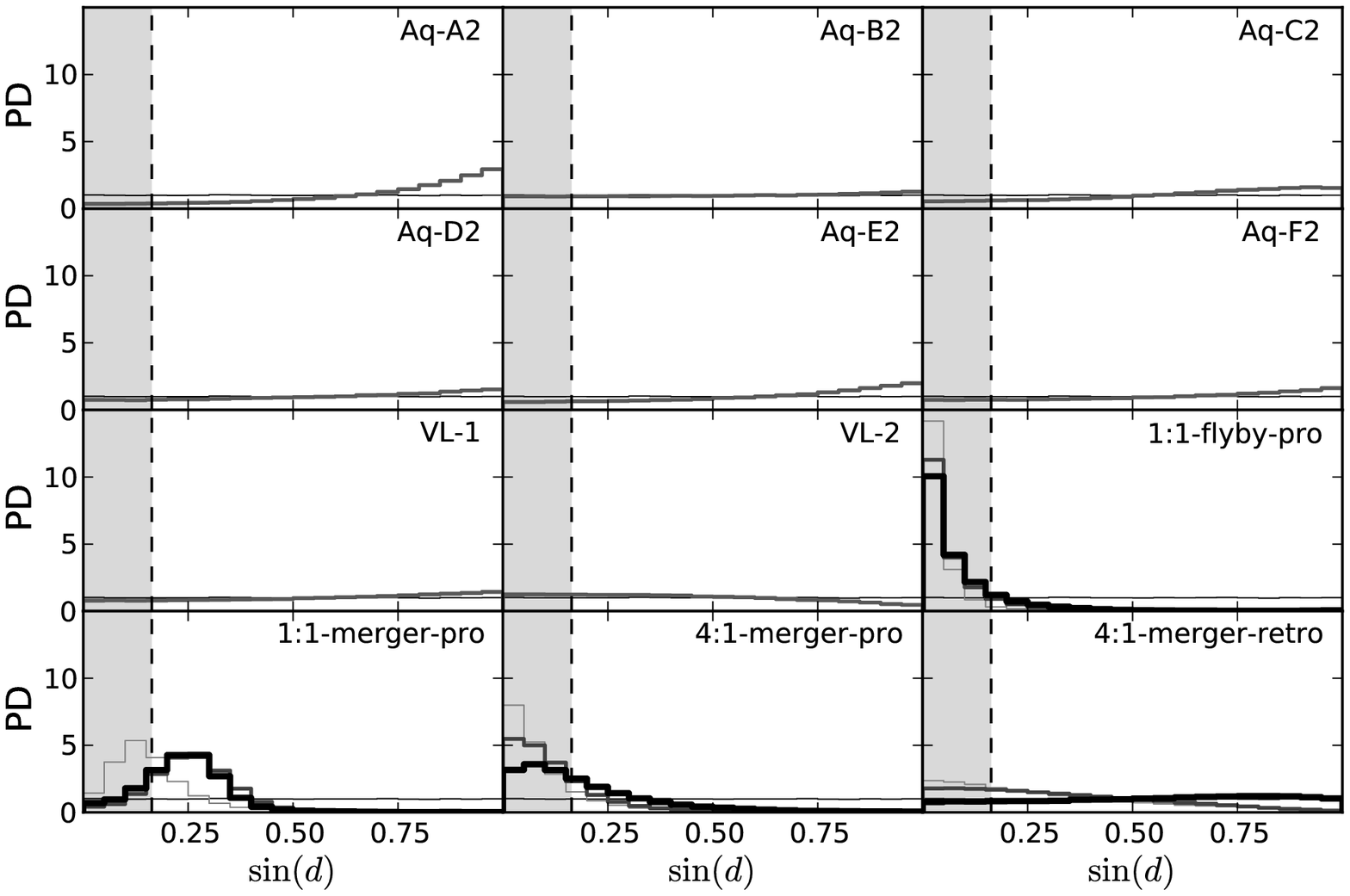}
 \caption{Distributions of the probability density (PD) of the angular distances of the average angular momentum vector from the equator for the 6 most closely angular momentum vectors out of 8 different, randomly drawn angular momentum vectors. Same panels as in Figure \ref{fig:standarddistance}. The vertical dashed line indicates the distance of the average orbital pole of the MW satellites from the equator of the MW, $d = 9.4^\circ$. To fulfill the orientation criterion, a realisation has to fall to the left of this line.
 As in Figure \ref{fig:lovell1}, the Aquarius and Via Lactea simulations produce results that make it unlikely to find a situation resembling the MW satellite orbital poles. Their subhalo angular momenta preferentially align with the main halos spin, which is perpendicular to the equator. An exception is the Via Lactea 2 simulations, which shows a slight excess at low $d$. All other cosmological models have a lower number of average angular momenta close to the equator than the isotropic case. Most models of galaxy interactions however naturally come up with average angular momenta of their tidal debris close to the equator, resembling the observed case of the MW satellites. 
 }
 \label{fig:averagedistance}
\end{figure*}

The distribution of angular distances $d$\ of the average angular momentum from the respective equator of the model are plotted in Figure \ref{fig:averagedistance}. The plot in fact shows $\sin(d)$, as this results in bins representing rings of equal area measured from the equator. This gives a flat distribution for the isotropic case and therefore eases the comparison of the various distributions.

All Aquarius simulations have their maximum probability density at the poles, far away from the equator. For the bins close to the equator their probability densities lie below the line corresponding to the isotropic case. They are therefore less likely to produce an average angular momentum vector close to the equator than the isotropic case. This results from the preferred alignment of subhalo orbits with the axis of the main halo spin, which has been illustrated in Figure \ref{fig:lovell1}. For model Aq-B2, its $\sin(d)$\ distribution is almost flat, mostly following the isotropic case, but with a minor increase towards the poles. 

The Via Lactea models again have a behaviour similar to that of the Aquarius models, however the VL-2 average angular momentum vectors do not peak close to the pole, but shows a small increase towards the equator compared to the isotropic case.

The tidal models again behave differently. Initially (5 Gyr), the $\sin(d)$-distributions of all four have their maximum close to the equator ($\sin(d) = 0$). They widen slightly for the models 1:1-flyby-pro and 4:1-merger-pro, the peaks become lower but do not move away from the equator much. This is different for model 1:1-merger-pro, in which the distribution peaks at $\sin(d) \approx 0.13$\ initially (5 Gyr), but moves to $\sin(d) \approx 0.25$\ ($d = 14.5^\circ$) thereafter (7.5 and 10 Gyr), indicating the precession of the angular momenta of the particles. 
The distributions of these three models drop to zero for $\sin(d) > 0.5$\ ($d > 30^\circ$) at all time steps. The directions of their average angular momentum vectors thus all align close to the equator of the models. This is different for model 4:1-merger-retro, which has $\sin(d)$-values spreading over the whole range. Its maximum close to the equator is the lowest initially and the $\sin(d)$-distribution of this model approaches the isotropic case for the last time step.

The orientation of the average angular momentum vector of the 6 best-aligned angular momentum vectors should be close to the equator in order to resemble the case of the polar orbits of the MW satellites, which have an angular distance of the average orbital pole from the MW equator of $9.4^\circ$, represented as dashed vertical lines in Figure \ref{fig:averagedistance}. To fulfil the \textit{orientation criterion}, the average angular distance from the equator has to be lower than or equal to this value. The fraction $f_{d}$\ fulfilling this criterion is calculated similar to $f_{\Delta}$.

The fractions of particles fulfilling the orientation criterion are compiled in Table \ref{tab:fractions}. In the isotropic case, 16 per cent of the realisations are close enough to the equator. As expected, all Aquarius simulations and Via Lactea 1 show lower fractions, by a factor of 1.1 (Aq-B2) up to 2.7 (Aq-A2) less. Only Via Lactea 2 gives a slightly higher fraction, by a factor of $1.2$.

The tidal models again give the highest fractions of realisations fulfilling the orientation criterion. In model 1:1-flyby-pro almost all (84-91 per cent) of the realisations lead to average angular momentum directions close to the equator. In model 1:1-merger-pro this is only 15-60 per cent, because the peak of its $d$ distribution moves further away from the equator for later time steps. For model 4:1-merger-pro the fraction is again high (53-83 per cent). In general, the fraction drops for later time steps (see also Tables \ref{tab:tidal4t1} and \ref{tab:tidal1t1}). The tidal model 4:1-merger-retro starts with a relatively high fraction at 5 Gyr (36 per cent), but this value drops quickly, such that at the final time step it is only 13 per cent, comparable to the cosmological models and lower than the isotropic case. Considering only particles having a distance of at least 60 kpc from the galactic centre gives a higher fraction of 29 per cent at the final time step. This is again at the lower end of the range of values for tidal models.

Doubling the accepted distance from the equator from $d_{\mathrm{MW}} = 9.4^\circ$\ to $d = 18.8^\circ$\ increases $f_{d}$\ in the merger case, leading to 75-92 (1:1-merger-pro) and 81-96 per cent (4:1-merger-retro). It about doubles $f_{d}$\ for the the Aquarius and Via Lactes simulations, the isotropic case and the 4:1-mass ratio retrograde merger model, and increases by up to 10 percentage points in the fly-by case.

\subsection{Combining the criteria}
\label{sect:combined}

To resemble the distribution of MW satellite orbital poles, \textit{both} criteria presented before have to be met \textit{simultaneously}. If the direction of the average angular momentum and the spherical standard distance would be independent, the combined fraction should simply be the product of the two fractions. If the underlying distribution of angular momentum directions is non-isotropic, this will not be the case any more. Both the actual fraction of realisations that fulfil both criteria, $f_{both}$, as well as the expected fraction if the two criteria were completely independent, $f_{indep}$, are compiled in Table \ref{tab:fractions}.

As can be expected, in the isotropic model $f_{\rm{both}}$\ and $f_{\rm{indep}}$\ are consistent with each other, 0.5 per cent of the realisations give parameters that fulfil both criteria together. It is highly unlikely to find the observed distribution of the MW satellite orbital poles if the satellite orbits were distributed isotropically around the MW.

It has been claimed by L11 that cosmological simulations naturally give rise to ''[...] distributions of coherently rotating satellites, such as those inferred in the Milky Way [...]''. We tested this claim by determining $f_{\rm{both}}$, finding it to be falsified. All six Aquarius models result in $f_{\rm{both}}$-values that are below the one for the isotropic case. The model displaying the strongest alignment of subhalo angular momenta with the main halo spin, Aq-A2, results in a MW-satellite-like distribution of orbital poles in only 0.1 per cent of the realisations. Averaging over the six Aquarius models gives a likelihood of 0.33 per cent. VL-1 is consistent with the Aquarius models, having $f_{\rm{both}} = 0.4$. Only VL-2 is more likely to reproduce the observed situation than the isotropic case, but still only in 1.5 per cent of the cases. The reason is that our crudely derived halo spin direction (the sum of angular momenta of a subset of particles from the simulation) does not point into the direction in which most subhalo angular momenta point.

Most tidal models, in contrast, result in much higher fractions passing both criteria. In the equal-mass fly-by model (1:1-flyby-pro), $f_{\rm{both}}$\ is as high as $60$\ per cent for the first time step. This value is only slightly reduced for the last time step, where it is $57$\ per cent. Therefore, this model will produce, insofar as we tested, angular momenta distributions with properties similar to that of the MW satellite orbital poles in most of the cases. The equal-mass merger case gives $f_{\rm{both}} = 48$\ per cent for the first time step (5 Gyr) and $12$\ per cent for the last (10 Gyr). While lower than in the fly-by case, this fraction is not low enough to rule out the model and is a factor of 27 larger than that of the best Aquarius simulation in this analysis and 8 times larger than the VL-2 result. The third tidal model (4:1-merger-pro) again gives very high $f_{\rm{both}}$-values, also dropping with time from $75$\ per cent at 5 Gyr to $46$\ per cent at 10 Gyr.

Only the last tidal model (4:1-merger-retro) shows low $f_{\rm{both}}$-values, dropping from 19 per cent (5 Gyr) to only 1 per cent (10 Gyr), a value still larger than (but only by a factor 2 to 3) the average of the cosmological models. If, however, angular momenta of the particles belonging to the spheroid-like component of the merger remnant are removed by demanding a minimum distance of 60 kpc from the central galaxy, then this model still has a $f_{\rm{both}}$\ of 13 per cent at the final time step.

As discussed in Sect. \ref{sect:orientation}, doubling the allowed distance from the equator increases $f_{d}$\ and consequently also increases $f_{\rm{both}}$. This leads to $f_{\rm{both}} = 67$\ per cent at the final time step for model 1:1-merger-pro and to $f_{\rm{both}} = 68$\ for model 4:1-merger-pro.
It only slightly changes $f_{\rm{both}}$\ for the other models, increasing the fraction by one percentage point in the fly-by model and doubling the fractions for the Aquarius and Via Lactea simulations, the isotropic case and the last tidal model (4:1-merger-retro).

Tables \ref{tab:tidal4t1} and \ref{tab:tidal1t1} in the Appendix compile the results for all tidal models of P11. High $f_{\rm{both}}$\ are common, some models even show values as high as 90 per cent or more. The $f_{\rm{both}}$-averages for the different model-types slightly decrease with time, but the fractions can increase for some models. Overall, they remain consistent for the different time steps. The $f_{\rm{both}}$-averages are highest in fly-by encounters (67 to 52 per cent for the investigated equal-mass and 71 to 41 per cent for the 4:1 mass-ratio models at time steps 5 and 10 Gyr, respectively). Retrograde mergers, on average, produce lower $f_{\rm{both}}$\ (31 to 33 per cent and 20 to 11 per cent) than prograde mergers (41 to 36 per cent and 44 to 27 per cent ). The statistical properties of the distributions of angular momentum directions in the tidal models are stable over many Gyr. In order to reproduce the observed orbital-pole distribution, it is not necessary to fine-tune the point in time after the galaxy-galaxy interaction. 

Taken together, our investigations show that tidal models are capable of naturally accounting for the observed distribution of MW satellite orbital poles. In contrast to that, it can most certainly not be claimed that the Aquarius or Via Lactea models naturally produce the phase-space distribution of the MW satellites. \textit{There is no evidence in those $\Lambda$CDM models that filamentary accretion can account for the peculiar properties of the MW satellite orbits.} Even an isotropic distribution of satellite angular momentum directions has a higher likelihood of accounting for the MW dSph phase-space correlation than all of the Aquarius simulations. 
Tidal models, however, can reproduce the observed properties with ease.

Are there structures in the Aquarius simulations that resemble the MW satellite system? 
An answer might be sought in the spacial distribution of the DM subhalos, as shown in figure 6 of L11. This figure depicts the positions of all subhalos, but only those are coloured which have orbital angular momenta aligned close to the axis of the main-halo spin ($\left|\cos (\Theta_{H \cdot S})\right| > 0.9$). As these subhalos all move within approximately the same plane, they form a 'quasi-planar' distribution. However, it must not be forgotten that halos on these orbits are only slightly over-abundant, at most by a factor of about 2.3 compared to an isotropic distribution of subhalo angular momentum directions (for Aq-A2, see Figure \ref{fig:lovell1}). The remaining subhalos are the majority and do not orbit in this plane. Thus, ignoring the suggestive colouring in figure 6 of L11, it is obvious that the spacial distribution of subhalos does not resemble that of a disc of satellites but is much wider. The 'quasi-planar' distribution of L11 is, in fact, only a subset of the rather 'quasi-spheroidal' distribution of DM subhalos in the Aquarius simulations. There is no mechanism to let luminous satellite galaxies only form in subhalos selected by their final position around the host halo they end up in. The chance to reproduce the observed, well pronounced DoS within the Aquarius models must therefore be extremely small. In addition, as discussed in Sections \ref{sect:assumptions}-\ref{sect:mpsatpoles}, the coloured subhalos of L11 emphasise a plane most likely oriented \textit{perpendicular} to the DoS.

\subsection{A different equator}
\label{sect:differentequator}

Relaxing the demand of the orientation criterion, one might consider the hypothetical situation that the disc galaxy axis in the DM halo does not align with the main halo spin. The most favourable situation for the cosmological simulations would then be the case in which the equator lies where most subhalo angular momentum directions cluster. Therefore, the equator can be defined as the great circle connecting the main halo spin direction and its anti-pole with the direction in which the highest density of subhalo angular momentum directions is found (see figure 4 in L11). Would the Aquarius models result in satisfactory $f_{\rm{both}}$-values in this case?

Even in this unlikely and extremely contrived situation for the cosmological models, their $f_{\rm{both}}$\ values can not compete with those of the tidal models. The highest value for $f_{\rm{both}}$, 3.6 per cent, is found for Aq-C2. The other models produce lower fractions: 2.7 (Aq-A2), 2.0 (Aq-E2), 1.7 (Aq-D2), 1.5 (Aq-F2) and 0.9 (Aq-B2) per cent. Averaged over the six models, the fraction is increased by only a factor of six, from 0.33 to 2.1 per cent.

\section{Conclusions}
\label{sect:concl}

A new method has been presented which estimates the likelihood of a given distribution of angular momentum directions to reproduce the peculiar properties of the observed MW satellite orbital poles, namely their preferred co-rotation on a near-polar orbit. For sets drawn from modelled angular momentum direction distributions, it tests both the closeness of the average direction of the six best-aligned angular momenta to the equator and their spherical standard distance. The method is easily applied to angular momentum distributions derived from any models and can be adjusted as more MW satellite galaxy orbital poles become available through observations of proper motions.

If the MW satellite orbital poles were drawn from the orbital angular momenta of $\Lambda$CDM subhalos determined from the Aquarius simulations, then the substantial simulation effort has not been able to arrive at a distribution with similar parameters as the one observed around the MW. It was found that for the most likely orientation of the disc galaxy spin, a similar alignment has a probability of at most 0.44 per cent to occur in the Aquarius simulations. This result does not agree with the suggestion put forward by L11 that coherently rotating quasi-planar distributions of satellites such as inferred observationally for the MW arise naturally in simulations of a $\Lambda$CDM universe. 

The L11 cosmological models preferentially lead to slight over-abundances of angular momenta of subhalos aligned with the axis of the main-halo spin and so most probably with the axis of rotation of the central galaxy. This alignment is in stark contrast to the polar orientation of the orbits of the MW satellites and the polar orientation of the Disc of Satellites. Even the highly unlikely and contrived case of a perpendicular orientation of main halo and MW disc spin increases the probability to find the observed orbital pole distribution by only a factor of 6 to about 2 per cent. 

An analysis of the two Via Lactea simulations gives similar results. With 1.5 per cent the VL-2 model has the highest probability, as the sub halo spin directions in this model do not preferentially align with our crudely derived direction of the main halo spin.

This result has consequences. \citet{Keller2012} analysed the distribution of globular clusters of the MW. Those globular clusters which are classified as young halo objects and are thought to have been stripped from accreted dwarf galaxies populate the same plane as the disc of satellite galaxies. \citet{Keller2012}, referring to the L11 results, have interpreted this as a sign of filamentary origin. They state: ''Simulations have shown that the planar arrangement of satellites can arise as filaments of the surrounding large scale structure feed into the Milky Way's potential. We therefore propose that our results are direct observational evidence for the accreted origin of the outer young halo globular cluster population''. In the light of our results, the opposite is the case. The addition of more objects to the DoS puts even stricter demands on the models trying to explain the observed situation. As filamentary accretion already fails at explaining the orbital poles of eight MW satellite galaxies, more objects distributed in the same planar structure (and therefore also orbiting in it, otherwise the structure would not be stable) only makes things worse.

In contrast to a cosmological origin of the MW satellites, a tidal origin can easily reproduce the observed parameters. This was tested with models of interacting galaxies from P11 having different mass ratios and resulting in both fly-by and merger cases. Most of these models not only naturally produce a clustering of the angular momenta, they also lead to discs of tidal debris around the central galaxy, similar to the DoS. Thus, a tidal interaction likely is the origin of the MW disc of satellites. The orientation of the orbital and spatial distribution of the tidal debris depend on the interaction geometry, which can well result in polar orbits and distributions. This is in agreement with the vast polar structure (VPOS) around the MW, which contains stellar and gaseous streams, globular clusters and satellite galaxies and their orbits (Pawlowski et al. 2012).

Indeed, other extra-galactic post-merger cases of aligned dSph satellites with tidal arms have emerged \citep{Malphrus1997, Bournaud2007, Galianni2010, Duc2011}, strengthening the notion that dSph and dE satellites are old TDGs \citep{Okazaki2000}. The implications of this for cosmological theory and fundamental physics is discussed in \citet{Kroupa2010, Kroupa2012}.

The percentages calculated in this work negate the suggestion that a satellite galaxy distribution like that of the MW arises \textit{naturally} in a $\Lambda$CDM universe. L11 argue that the six Aquarius halos can be considered to be approximately representative of the population of MW-sized halos as a whole. The similar results for the independent Via Lactea models seem to support this claim. Therefore, it can be expected that further simulations in the $\Lambda$CDM framework will not lead to significantly different results.
Since the structure formation and merger history is similar in a warm-dark-matter cosmology (only the number of subhalos drops), the same conclusion holds true there as well. \citet{Knebe2008} have compared the anisotropic spatial distribution of DM subhalos within their host halos for cold dark matter (CDM) and warm dark matter (WDM) simulations. They state that the spatial anisotropy of subhalos in the WDM model is in fact \textit{less} pronounced than in the CDM case.
Thus, CDM as well as WDM models appear to be ruled out.

The tidal scenario, in which the MW satellite galaxies are TDGs formed in a galaxy-encounter, is a promising alternative for their origin and able to explain the phase-space-correlation of the DoS and VPOS \citep{Pawlowski2011,Pawlowski2012}.

\section*{Acknowledgments}
M.S.P. and G.H. acknowledge support through DFG reseach grant KR1635/18-2 and HE1487/36-2, respectively, in the frame of the \textit{DFG Priority Programme 1177, “Witnesses of Cosmic History: Formation and evolution of galaxies, black holes, and their environment”}. M.S.P. also acknowledges support of the \textit{Bonn-Cologne Graduate School of Physics and Astronomy}. B. F. acknowledges the support of the \textit{Alexander von Humboldt-Foundation}.
We thank Mark R. Lovell for providing the angular momentum direction vectors he derived from the Aquarius simulations. We also thank Manuel Metz for providing the orbital pole data of the MW satellite galaxies.

\bibliographystyle{mn2e}
\bibliography{orbitalpoles}

\appendix

\section{Tables for tidal models}

Tables \ref{tab:tidal4t1} and \ref{tab:tidal1t1} compile the results of analysing three different time steps of all 72 tidal-interaction models of P11 with the method presented in this paper. See P11 for detailed descriptions of the model set-up. The 4:1 mass ratio models have a target galaxy more massive than the infalling galaxy. In all encounters, the infalling galaxy is oriented perpendicular to the target galaxy and approaches it on a polar orbit. The model names consist of two parts. The first, (e.g. '5deg') describes the angle between the velocity vector and the position vector of the two galaxies in degrees (here $5^\circ$), larger numbers give larger perigalactica. The second part (e.g. '100vel') describes the initial relative velocity of the two galaxies with respect to the parabolic velocity $v_{\mathrm{parab}}$\ in per cent (here $1.0~v_{\mathrm{parab}}$). The fractions compiled in the tables have been calculated from $10^4$\ realisations. Assuming Poission statistics, the uncertainties $\Delta f$\ can be calculated from the fractions $f$\ as $\Delta f = \sqrt{f} / 10$, where both $\Delta f$\ and $f$\ are given in per cent.

\begin{table*}
\begin{minipage}{180mm}
 \caption{4:1 mass ratio tidal models and resulting fractions of realisations fulfilling the criteria}
 \label{tab:tidal4t1}
 \begin{center}
 \begin{tabular}{@{}llcccccccccccc}
  \hline
  
 & & \multicolumn{4}{c}{Time step 5 Gyr} & \multicolumn{4}{c}{Time step 7.5 Gyr} & \multicolumn{4}{c}{Time step 10 Gyr} \\  
  Type & Model & $N$ & $f_{\Delta} [\%]$ & $f_{d} [\%]$ & $f_{\rm{both}} [\%]$ & $N$ & $f_{\Delta} [\%]$ & $f_{d} [\%]$ & $f_{\rm{both}} [\%]$ & $N$ & $f_{\Delta} [\%]$ & $f_{d} [\%]$ & $f_{\rm{both}} [\%]$\\
  \hline

fb & 2deg200vel & 1045 & 99.4 & 99.9 & \it 99.3 & 1136 & 98.4 & 99.3 & \it 98.0 & 1084 & 98.3 & 98.1 & \it 97.0 \\
fb & 4deg175vel & 11882 & 66.0 & 79.9 & \it 59.8 & 12389 & 29.5 & 63.6 & \it 26.5 & 11793 & 31.8 & 49.8 & \it 24.7 \\
fb & 4deg200vel & 14127 & 34.7 & 77.6 & \it 33.8 & 13732 & 36.3 & 70.8 & \it 33.2 & 12205 & 39.9 & 69.2 & \it 36.4 \\
fb & 4deg225vel & 16477 & 28.8 & 83.9 & \it 28.6 & 16812 & 27.4 & 71.0 & \it 25.7 & 16276 & 25.6 & 56.9 & \it 21.0 \\
fb & 4deg250vel & 11858 & 38.3 & 87.0 & \it 38.1 & 12016 & 48.9 & 92.2 & \it 48.9 & 11330 & 45.0 & 85.3 & \it 44.4 \\
fb & 6deg175vel & 6284 & 97.6 & 86.7 & \it 85.6 & 7730 & 42.3 & 67.5 & \it 37.1 & 8245 & 27.4 & 41.9 & \it 18.3 \\
fb & 6deg200vel & 6205 & 97.9 & 95.2 & \it 93.7 & 7730 & 38.8 & 65.0 & \it 33.0 & 7337 & 31.5 & 56.4 & \it 24.9 \\
fb & 6deg225vel & 8659 & 98.7 & 94.7 & \it 93.8 & 8213 & 85.4 & 82.1 & \it 73.5 & 8509 & 57.4 & 60.1 & \it 40.7 \\
fb & 6deg250vel & 4970 & 94.3 & 96.0 & \it 91.4 & 5067 & 77.4 & 62.0 & \it 50.3 & \multicolumn{4}{c}{no data, galaxies left model volume}  \\
fb & 8deg175vel & 4522 & 96.2 & 93.1 & \it 91.0 & 5129 & 80.9 & 84.3 & \it 73.5 & 4815 & 74.2 & 76.5 & \it 63.4 \\
\multicolumn{2}{l}{Average} & & 75.2 & 89.4 & \it 71.5 & & 56.5 & 75.8 & \it 50.0 & & 47.9 & 66.0 & \it 41.2 \\
      \hline

mp & 2.5deg050vel & 66917 & 32.8 & 12.7 & \it 6.9 & 62262 & 23.9 & 27.4 & \it 10.8 & 63893 & 14.8 & 28.2 & \it 6.2 \\
mp & 2.5deg100vel & 81684 & 44.9 & 23.9 & \it 16.2 & 82953 & 31.2 & 17.7 & \it 7.4 & 81788 & 25.9 & 27.2 & \it 7.5 \\
mp & 5deg050vel & 64938 & 38.4 & 24.9 & \it 16.6 & 60115 & 29.0 & 17.8 & \it 7.9 & 64109 & 24.3 & 28.9 & \it 6.9 \\
mp & 5deg100vel & 86824 & 86.1 & 68.5 & \it 63.5 & 82448 & 77.8 & 58.5 & \it 51.1 & 81435 & 75.9 & 59.9 & \it 50.4 \\
mp & 7.5deg100vel & 93940 & 82.6 & 82.3 & \it 74.6 & 80454 & 89.1 & 74.6 & \it 70.9 & 77768 & 74.0 & 53.9 & \it 47.0 \\
mp & 10deg100vel & 83850 & 89.8 & 90.0 & \it 85.7 & 72029 & 77.1 & 74.3 & \it 64.0 & 68751 & 65.2 & 53.2 & \it 43.4 \\
\multicolumn{2}{l}{Average} & & 62.4 & 50.4 & \it 43.9 & & 54.7 & 45.0 & \it 35.3 & & 46.7 & 41.9 & \it 26.9 \\
    \hline

mr & 2.5deg050vel & 69336 & 27.8 & 72.8 & \it 25.9 & 65802 & 25.7 & 45.9 & \it 18.3 & 72171 & 25.7 & 43.1 & \it 17.2 \\
mr & 2.5deg100vel & 80994 & 42.1 & 53.0 & \it 29.1 & 78535 & 21.6 & 41.7 & \it 13.4 & 78556 & 22.9 & 45.2 & \it 15.5 \\
mr & 5deg050vel & 64109 & 30.8 & 62.8 & \it 25.3 & 61710 & 25.3 & 46.7 & \it 16.9 & 64159 & 20.9 & 45.3 & \it 14.5 \\
mr & 5deg100vel & 72679 & 22.9 & 42.8 & \it 15.7 & 61278 & 19.7 & 34.2 & \it 10.6 & 59795 & 23.1 & 41.1 & \it 15.3 \\
mr & 7.5deg100vel & 70490 & 56.5 & 4.3 & \it 3.3 & 51645 & 41.1 & 36.3 & \it 18.4 & 49391 & 17.4 & 19.4 & \it 4.4 \\
mr & 10deg100vel & 75431 & 32.8 & 36.2 & \it 18.8 & 60248 & 16.2 & 28.4 & \it 6.9 & 61157 & 7.9 & 13.8 & \it 0.9 \\
\multicolumn{2}{l}{Average} & & 35.5 & 45.3 & \it 19.7 & & 24.9 & 38.9 & \it 14.1 & & 19.6 & 34.6 & \it 11.3 \\

  \hline
 \end{tabular}
 \end{center}

 \small \smallskip

Type: Type of interaction. Either a fly-by with the infalling disc in prograde orientation (fb), or a merger with the infalling disc in prograde (mp) or retrograde (mr) orientation.
Model: Name of the model or simulation.
To show the evolution of the models, the following data is given for three different time steps in each model, except when stated otherwise.
$N$: Number of angular momentum directions to draw realisations from.
$f_{\Delta}$: Fraction of realisations fulfilling the clustering criterion (having a spherical standard distance of no more then $35.4^\circ$, the value for the MW satellites). 
$f_{d}$: Fraction of realisations fulfilling the direction criterion (having an average angular momentum direction pointing no further away from the equator than $9.4^\circ$, the value for the MW satellites).
$f_{\rm{both}}$: Fraction of realisations fulfilling both criteria. They are printed in \textit{italic} to ease the comparison of the different columns.
\end{minipage}
\end{table*}

\begin{table*}
\begin{minipage}{180mm}
 \caption{Equal-mass tidal models and resulting fractions of realisations fulfilling the criteria}
 \label{tab:tidal1t1}
 \begin{center}
 \begin{tabular}{@{}llcccccccccccc}
  \hline
  
 & & \multicolumn{4}{c}{Time step 5 Gyr} & \multicolumn{4}{c}{Time step 7.5 Gyr} & \multicolumn{4}{c}{Time step 10 Gyr} \\  
  Type & Model & $N$ & $f_{\Delta} [\%]$ & $f_{d} [\%]$ & $f_{\rm{both}} [\%]$ & $N$ & $f_{\Delta} [\%]$ & $f_{d} [\%]$ & $f_{\rm{both}} [\%]$ & $N$ & $f_{\Delta} [\%]$ & $f_{d} [\%]$ & $f_{\rm{both}} [\%]$\\
  \hline

fb & 5deg180vel & 3279 & 86.5 & 96.6 & \it 85.8 & 3391 & 87.0 & 94.9 & \it 85.4 & 3355 & 82.2 & 89.2 & \it 78.1 \\
fb & 5deg200vel & 5821 & 59.4 & 91.3 & \it 59.4 & 5841 & 59.6 & 86.6 & \it 58.9 & 5756 & 58.7 & 83.4 & \it 57.5 \\
fb & 5deg220vel & 3991 & 46.5 & 90.4 & \it 46.5 & 3800 & 55.0 & 89.9 & \it 54.9 & 3642 & 44.8 & 82.0 & \it 44.3 \\
fb & 5deg240vel & 2189 & 76.5 & 99.2 & \it 76.5 & 1871 & 71.2 & 97.3 & \it 71.2 & 1596 & 61.9 & 92.3 & \it 61.8 \\
fb & 6deg180vel & 2108 & 97.7 & 99.1 & \it 97.3 & 1998 & 93.7 & 98.9 & \it 93.4 & 2125 & 86.0 & 95.6 & \it 85.1 \\
fb & 6deg200vel & 4701 & 76.3 & 93.5 & \it 75.6 & 4671 & 56.1 & 81.1 & \it 54.1 & 4793 & 44.1 & 70.4 & \it 41.3 \\
fb & 6deg220vel & 3732 & 45.8 & 91.8 & \it 45.8 & 3840 & 40.9 & 82.0 & \it 40.6 & 3805 & 32.9 & 71.8 & \it 31.5 \\
fb & 6deg240vel & 3797 & 31.0 & 95.2 & \it 31.0 & 3464 & 32.7 & 92.8 & \it 32.7 & 3093 & 37.2 & 84.4 & \it 36.7 \\
fb & 7deg180vel & 1636 & 98.3 & 99.7 & \it 98.2 & 1712 & 93.6 & 97.1 & \it 92.5 & 1959 & 65.1 & 70.5 & \it 57.8 \\
fb & 7deg200vel & 3010 & 68.0 & 94.5 & \it 67.8 & 3361 & 48.2 & 77.7 & \it 45.8 & 3297 & 42.3 & 67.4 & \it 37.8 \\
fb & 7deg220vel & 2209 & 50.5 & 95.6 & \it 50.5 & 2406 & 35.2 & 80.6 & \it 34.7 & 2277 & 31.1 & 71.8 & \it 29.2 \\
fb & 7deg240vel & 2423 & 39.8 & 94.2 & \it 39.8 & 2212 & 45.2 & 94.6 & \it 45.1 & 2125 & 57.5 & 94.5 & \it 57.3 \\
fb & 8deg180vel & 1353 & 99.1 & 99.2 & \it 98.6 & 1433 & 95.5 & 97.9 & \it 94.6 & 1666 & 75.7 & 82.3 & \it 69.9 \\
fb & 8deg200vel & 1939 & 82.5 & 98.4 & \it 82.4 & 2077 & 69.1 & 88.8 & \it 66.7 & 2232 & 43.6 & 72.0 & \it 40.6 \\
fb & 8deg220vel & 1257 & 66.2 & 96.2 & \it 66.0 & 1292 & 63.8 & 88.6 & \it 62.5 & 1214 & 51.3 & 75.6 & \it 47.3 \\
fb & 8deg240vel & 1298 & 42.3 & 92.1 & \it 42.3 & 1190 & 52.3 & 96.3 & \it 52.2 & 1112 & 61.9 & 93.9 & \it 60.9 \\
\multicolumn{2}{l}{Average} & & 66.6 & 95.4 & \it 66.5 & & 62.4 & 90.3 & \it 61.6 & & 54.8 & 81.1 & \it 52.3 \\
       \hline

mp & 0deg050vel & 11212 & 53.1 & 2.3 & \it 2.1 & 16610 & 42.3 & 34.0 & \it 12.1 & 17331 & 56.9 & 53.0 & \it 33.8 \\
mp & 0deg100vel & 14095 & 51.3 & 20.8 & \it 18.0 & 15821 & 51.4 & 11.9 & \it 6.3 & 16217 & 43.6 & 51.9 & \it 27.9 \\
mp & 2.5deg050vel & 11452 & 46.7 & 1.4 & \it 0.8 & 17368 & 39.3 & 27.6 & \it 7.6 & 14000 & 45.5 & 37.9 & \it 17.3 \\
mp & 2.5deg100vel & 11353 & 30.3 & 19.4 & \it 10.0 & 11652 & 28.2 & 16.9 & \it 2.9 & 12825 & 30.9 & 36.3 & \it 12.6 \\
mp & 2.5deg150vel & 16308 & 35.5 & 55.6 & \it 26.9 & 14332 & 28.0 & 25.3 & \it 8.5 & 14888 & 35.3 & 22.3 & \it 5.8 \\
mp & 5deg050vel & 13122 & 68.1 & 8.2 & \it 6.1 & 13986 & 68.6 & 14.4 & \it 4.9 & 13581 & 54.7 & 33.2 & \it 17.4 \\
mp & 5deg100vel & 19932 & 76.5 & 13.4 & \it 9.5 & 19894 & 76.1 & 2.5 & \it 0.3 & 20754 & 78.5 & 21.0 & \it 14.3 \\
mp & 5deg150vel & 32908 & 57.6 & 71.7 & \it 52.5 & 32715 & 51.6 & 53.2 & \it 34.0 & 29580 & 43.7 & 29.4 & \it 11.9 \\
mp & 7.5deg050vel & 14772 & 72.2 & 38.3 & \it 29.9 & 14874 & 84.5 & 9.3 & \it 4.4 & 14414 & 79.4 & 29.1 & \it 20.9 \\
mp & 7.5deg100vel & 36438 & 74.4 & 59.2 & \it 48.4 & 35954 & 78.4 & 15.1 & \it 8.7 & 37302 & 77.5 & 19.7 & \it 11.5 \\
mp & 7.5deg150vel & \multicolumn{4}{c}{not completely merged} & 33787 & 74.3 & 66.8 & \it 57.8 & 34102 & 68.9 & 33.4 & \it 24.7 \\
mp & 10deg050vel & 18491 & 68.1 & 65.2 & \it 50.5 & 21126 & 90.9 & 5.9 & \it 3.9 & 19152 & 90.4 & 11.0 & \it 7.8 \\
mp & 10deg100vel & 30105 & 95.7 & 95.5 & \it 93.5 & 28464 & 95.9 & 68.0 & \it 65.8 & 30895 & 96.2 & 34.2 & \it 32.6 \\
mp & 15deg050vel & 26512 & 82.1 & 85.6 & \it 76.9 & 29233 & 94.1 & 76.2 & \it 72.6 & 28527 & 92.6 & 92.5 & \it 88.9 \\
mp & 15deg100vel & 22397 & 100.0 & 100.0 & \it 100.0 & 23707 & 100.0 & 99.4 & \it 99.4 & 25586 & 100.0 & 93.3 & \it 93.3 \\
mp & 20deg050vel & 31111 & 99.9 & 99.7 & \it 99.6 & 26462 & 99.8 & 89.1 & \it 88.9 & 30140 & 99.6 & 87.7 & \it 87.6 \\
mp & 20deg100vel & \multicolumn{4}{c}{not completely merged} & 31581 & 100.0 & 100.0 & \it 100.0 & 33832 & 100.0 & 99.6 & \it 99.6 \\
\multicolumn{2}{l}{Average} & & 67.4 & 49.1 & \it 41.7 & & 70.8 & 42.1 & \it 34.0 & & 70.2 & 46.2 & \it 35.8 \\
     \hline
     
mr & 0deg050vel & 11495 & 53.5 & 1.8 & \it 1.6 & 16130 & 44.7 & 27.4 & \it 10.0 & 18982 & 60.2 & 55.7 & \it 38.0 \\
mr & 0deg100vel & 14022 & 53.2 & 21.9 & \it 19.5 & 14857 & 50.4 & 8.8 & \it 4.4 & 16146 & 44.9 & 52.5 & \it 29.2 \\
mr & 2.5deg050vel & 12295 & 72.9 & 20.9 & \it 20.7 & 14310 & 63.0 & 48.2 & \it 34.9 & 14885 & 61.0 & 66.6 & \it 48.0 \\
mr & 2.5deg100vel & 17857 & 48.7 & 27.9 & \it 25.6 & 18968 & 67.4 & 23.9 & \it 19.8 & 19318 & 56.5 & 50.4 & \it 34.3 \\
mr & 2.5deg150vel & 15386 & 86.9 & 84.7 & \it 78.9 & 14743 & 90.1 & 61.5 & \it 58.2 & 16047 & 90.8 & 74.9 & \it 70.2 \\
mr & 5deg050vel & 13029 & 70.4 & 28.8 & \it 28.4 & 17994 & 67.7 & 63.8 & \it 49.9 & 16502 & 56.7 & 54.5 & \it 36.4 \\
mr & 5deg100vel & 18057 & 40.7 & 19.3 & \it 17.0 & 19634 & 46.5 & 48.1 & \it 31.8 & 17601 & 48.4 & 43.4 & \it 26.7 \\
mr & 5deg150vel & 13071 & 99.1 & 98.8 & \it 98.1 & 15452 & 98.1 & 48.0 & \it 47.5 & 12727 & 98.3 & 38.1 & \it 37.4 \\
mr & 7.5deg050vel & 14598 & 82.6 & 63.7 & \it 61.8 & 18786 & 64.1 & 70.5 & \it 54.4 & 17718 & 80.6 & 41.5 & \it 34.7 \\
mr & 7.5deg100vel & 17127 & 82.7 & 1.6 & \it 0.5 & 17548 & 80.2 & 19.4 & \it 12.8 & 17102 & 71.4 & 52.9 & \it 43.4 \\
mr & 7.5deg150vel & \multicolumn{4}{c}{not completely merged} & 14666 & 34.5 & 22.7 & \it 5.7 & 13710 & 53.5 & 26.4 & \it 16.3 \\
mr & 10deg050vel & 11975 & 88.4 & 81.8 & \it 76.5 & 13713 & 79.4 & 83.3 & \it 73.9 & 15931 & 83.7 & 39.9 & \it 34.6 \\
mr & 10deg100vel & 11233 & 59.7 & 8.4 & \it 1.5 & 13193 & 80.2 & 11.2 & \it 8.6 & 12837 & 87.2 & 55.6 & \it 50.5 \\
mr & 15deg050vel & 11961 & 79.1 & 23.2 & \it 20.1 & 10837 & 85.4 & 60.5 & \it 55.5 & 10105 & 87.1 & 7.2 & \it 5.1 \\
mr & 15deg100vel & 5163 & 25.0 & 29.4 & \it 12.0 & 5762 & 20.8 & 28.2 & \it 8.6 & 6115 & 21.5 & 40.4 & \it 13.4 \\
mr & 20deg050vel & 11391 & 82.2 & 5.1 & \it 3.0 & 12510 & 89.7 & 35.8 & \it 33.1 & 10977 & 86.4 & 8.9 & \it 6.7 \\
mr & 20deg100vel & \multicolumn{4}{c}{not completely merged} & 9522 & 84.2 & 96.5 & \it 83.3 & 7745 & 67.5 & 44.5 & \it 33.9 \\
\multicolumn{2}{l}{Average} & & 68.3 & 34.5 & \it 31.0 & & 67.4 & 44.6 & \it 34.8 & & 68.0 & 44.3 & \it 32.9 \\

  \hline
 \end{tabular}
 \end{center}

 \small \smallskip

Labels as in Table \ref{tab:tidal4t1}.
\end{minipage}
\end{table*}

\label{lastpage}

\end{document}